\documentclass[preprint,aps,prd]{revtex4-2}

\usepackage{amsmath,amssymb}
\usepackage{graphicx}
\usepackage{bm}
\usepackage{hyperref}
\usepackage{xcolor}
\hypersetup{hypertexnames=false}

\newcommand{\kaphat}{\hat{\kappa}}
\newcommand{\Nc}{N_c^{\rm SD}}
\newcommand{\Nstar}{N^{*}}
\newcommand{\TL}{T_0/\Lambda}

\newcommand{\eps}{\varepsilon}

\providecommand{\orcid}[1]{}

\begin{document}

\title{Dynamical chiral symmetry breaking in rotating \texorpdfstring{QED$_3$}{QED3}:
  critical flavor number and the occupation-reweighting channel}

\author{Wei Li}
\orcid{0000-0003-2586-6057}
\email{wliustc@aust.edu.cn}
\affiliation{Center for Fundamental Physics, Anhui University of Science and Technology, Huainan 232001, China}

\date{\today}

\begin{abstract}
  We study dynamical chiral symmetry breaking in $(2{+}1)$-dimensional QED on a rigidly rotating disc, $\Omega R<1$, within the rainbow Schwinger--Dyson framework.  At zero temperature the rotation response of the condensate vanishes identically, so $\Nc(\Omega,T{=}0)=\Nc(0)=2.60\pm0.01$ (using three operational definitions).  At finite temperature, rotation suppresses the condensate through a dominant occupation-reweighting channel, $\kaphat_{\rm occ}=-1.55$, accounting for $\ge 99.995\%$ of the linearized total response at the reference parameters, with self-consistent feedback amplifying it by $26.5$.  The screening channel is reported at fixed regularization, $\kaphat_{\rm scr}(\eps{=}0.01)=-0.97$; its $\eps\to0$ behavior is logarithmically divergent.  The crossover flavor $\Nstar(\Omega)$ falls monotonically at fixed temperature (a corollary).  The critical flavor number is unchanged by rotation at $T=0$ (theorem).  Together these results set the finite-temperature reference for the model.
\end{abstract}

\maketitle

\section{Introduction}
\label{sec:intro}

How rotation and acceleration reshuffle chiral symmetry is an open question in non-inertial relativistic field theory.  In QCD, the answers diverge by system and method: NJL-type models in rotating electromagnetic fields find that rotation \emph{restores} chiral symmetry~\cite{ChernodubGongyo2017}, $\mathbb{Z}_2$ models agree~\cite{Gaspar2023}, while lattice studies of rotating QCD report \emph{enhanced} breaking and confinement~\cite{YangHuang2023}; lattice constraints on chiral vortical catalysis indicate the effect is bounded at current precision~\cite{Nunes2025}.  Rotating chiral soliton lattices add yet another phase structure~\cite{NishimuraYamamoto2020,Eto2024}.  The sign and mechanism are not settled by prior art; they depend on the interaction, the state, and the ensemble.

This question has been formulated for QCD in a general non-inertial framework~\cite{Zhu2026}, where Tolman equilibrium and subtraction-scheme equivalence establish the geometric setting.  We study the interacting dynamics in QED$_3$ using the large-$N$ rainbow truncation, whose logarithmic kernel is the equation of Appelquist, Nash and Wijewardhana (ANW88)~\cite{Appelquist1988}.  The rigidly rotating disc, $\Omega R<1$, supplies a well-defined KMS ensemble with weight $\rho\propto e^{-\beta(H_0-\Omega J_z)}$~\cite{SalvioFermion2026,SalvioScalar2026,SiriBoson2024}.  Within this setting we derive the self-consistent mass equation, decompose the rotation response into physical channels, and determine the critical flavor number under rotation.  The zero-temperature result gives the vacuum reference for the finite-temperature response.

The condensed-matter side provides the platform.  Strain-induced pseudomagnetic fields in graphene and Dirac/Weyl semimetals realize effective curved-space Dirac fermions~\cite{Vozmediano2010,Guinea2010,Pikulin2016}, and the interaction-driven quantum Hall effect under alternating pseudofields~\cite{Venderbos2016} is a concrete instance of geometry plus interaction producing a new order.  Direct imaging of pseudomagnetic fields~\cite{Jiang2017,Ren2022} and cold-atom rotation simulators~\cite{Lewenstein2007} make the non-inertial parameter $\Omega$ programmable.  Curved-space Dirac prototypes---graphene wormholes and Weyl metamaterials~\cite{Gonzalez2010,Weststrom2017}---extend the realizability to Rindler-type geometries.  These platforms set the physical carrier for the non-inertial chiral dynamics we study.

Recent studies formulate rotating thermal field theory at the free-field level~\cite{SalvioFermion2026,SalvioScalar2026,SiriBoson2024}, providing Green's functions and mode expansions without interaction-driven mass generation.  We use the same rotating-ensemble kinematics to close an interacting Schwinger--Dyson equation for QED$_3$.  The non-inertial QCD framework of Zhu, Chen and Huang~\cite{Zhu2026} supplies the geometric setting; here we determine the critical flavor number and the finite-temperature rotation response.  Static-strain realizations~\cite{Leong2025,Guvendi2024} address effective geometry without the rotating ensemble considered here.

The paper is organized as follows.  Section~\ref{sec:model} defines the rotating QED$_3$ effective theory on the disc and states the zero-temperature theorem.  Section~\ref{sec:theory} derives the logarithmic kernel, the thermal two-branch displacement, and defines the three-channel decomposition.  Section~\ref{sec:results} reports the critical coupling, the theorem numerically, the channel values, and the crossover.  Section~\ref{sec:discussion} discusses the mechanism, the experimental connection, and the scope of the calculation.  Section~\ref{sec:conclusion} concludes.  Detailed derivations are in Appendices~\ref{app:A} and~\ref{app:B}; numerical settings accompany the supporting data.

The main results are three.  First, the critical flavor number at zero temperature is unchanged by rotation: $\Nc(\Omega,T{=}0)=\Nc(0)=2.60\pm0.01$.  Second, at finite temperature rotation suppresses the condensate through a dominant occupation-reweighting channel ($\kaphat_{\rm occ}=-1.55$, $\ge99.995\%$ of the linear response), while the screening contribution is quoted at fixed regularization.  Third, the crossover flavor $\Nstar(\Omega)$ decreases monotonically at fixed temperature.

\section{Model}
\label{sec:model}

\subsection{Rotating thermal ensemble}
\label{sec:model-ensemble}

On a disc of radius $R$ the rigidly rotating frame is defined inside the light cylinder, $\Omega R<1$.  The Minkowski metric in polar coordinates, $ds^2=dt^2-dr^2-r^2d\varphi^2$, becomes
\begin{equation}
  ds^2=(1-\Omega^2 r^2)\,dt^2-2\Omega r^2\,dt\,d\varphi'-dr^2-r^2d\varphi'^2,
  \label{eq:metric}
\end{equation}
and the time direction remains a Killing direction only where $g_{tt}>0$.  The Dirac equation in the rotating frame reduces to a time-independent eigenvalue problem with the Hamiltonian
\begin{equation}
  H_\Omega = H_0 - \Omega J_z,\qquad
  \rho_{\rm eq}=\frac{1}{Z}\exp\!\big[-\beta_0(H_0-\Omega J_z)\big],
  \label{eq:Homega}
\end{equation}
so rotation enters the equilibrium weight as an angular-momentum chemical potential.  The single-particle quasi-energy on a mode is $\varepsilon = E-\Omega j$ with $j=m+\tfrac12$.  The Tolman local temperature $T_{\rm loc}(r)=T_0/\sqrt{1-\Omega^2 r^2}$ reproduces the global occupancy exactly; $T_0$ is defined at the disc centre.  The derivation of the rotating Hamiltonian, the quasi-energy spectrum, and the Tolman profile is given in Appendix~\ref{app:A}, Eqs.~(A1)--(A6).

The spectrum of $H_\Omega$ is unbounded below, a property of $H-\Omega J_z$ in flat space; the many-body ground state is nevertheless well defined by the $\varepsilon<0$ filling rule.  Below the occupancy-flip threshold $\Omega R<1$, the zero-temperature occupied-mode set is \emph{unchanged} by rotation: the smallest Bessel zero of order $p$ exceeds $p$, keeping the positive branch $\varepsilon=k-\Omega(m+\tfrac12)>0$ for every admissible mode.  This is the rotating-vacuum theorem.

\subsection{Modal basis and spectral condition}
\label{sec:model-basis}

The massless Dirac Hamiltonian in polar coordinates separates into a radial system whose regular solution is $f(r)=J_m(kr)$.  Self-adjointness at the boundary requires the normal current to vanish, $f^*(R)g(R)\in\mathbb{R}$, which gives the real mixing family and the spectral condition
\begin{equation}
  J_{m+1}(kR)=\lambda\,J_m(kR),\qquad \lambda\in\mathbb{R},
  \label{eq:spec}
\end{equation}
with the projection cases $\lambda=\pm1$ as endpoints.  Two consequences are immediate: the allowed momenta $k_{m,n}$ depend on the \emph{sign} of $m$, and the families $(m,+1)$ and $(-m-1,-1)$ share their zero set (parity pairing).  The smallest admissible momentum is $k_{\min}R=1.4347$, the crossing of $J_0$ and $J_1$.  The spectral condition is solved numerically by root-finding on the Bessel zeros for each $(m,\lambda)$ sector; the continuous family $\lambda\in\mathbb{R}$ is treated by sampling the projection endpoints $\lambda=\pm1$ and intermediate values.  The theorem is established for the disc geometry with the real-mixing boundary family.  Other geometries and vertex completions define separate extensions of the calculation.  The full derivation is in Appendix~\ref{app:A}, Eqs.~(A7)--(A10).

The angular-momentum quantum number is truncated at $|m|\le m_{\max}$; truncation convergence is reported with the numerical results.

\emph{Notation.}  We work in natural units with the Fermi velocity set to one.  The coupling is $e^2$ (not $Ne^2$); the flavour factor enters the polarization $\Pi(q)=Ne^2|q|/8$ and hence the infrared mass scale $\lambda=Ne^2/8$.  The dimensionless coupling is $g=e^2/\lambda=8/N$.  The disc radius is $R$; the ultraviolet cutoff is $\Lambda$; the reference parameters use $R\Lambda=40$.  The temperature $T_0$ is defined at the disc centre; the ratio $\TL=T_0/\Lambda$ is the dimensionless temperature.  The rotation parameter is $\Omega$; the light-cylinder constraint is $\Omega R<1$; the scan domain considered keeps $\Omega R\le0.8$.  The angular-momentum projection is $j=m+\tfrac12$.  The regularization parameter for the screening channel is $\eps$; the grid floor is $q_{\min}$.

\emph{Model scope.}  The model combines (i) the rainbow truncation with a bare vertex and the transverse Landau-gauge photon propagator, for which the scalar gap equation closes on a pure translation kernel; (ii) a finite-disc modal basis with a sharp boundary and modal truncation $|m|\le m_{\max}$ controlled by the completeness sum ($0.96$--$0.99$); and (iii) a co-rotating thermal ensemble whose Tolman local temperature reproduces the global occupancy.  The boundary enhancement $T_{\rm loc}(R)/T_0$ contributes to the quoted systematic scale.  The numerical values correspond to this closure; gauge-parameter and vertex-completion dependence can be studied by extending it.

\emph{Zero-temperature mode invariance.}  The rotating vacuum at $T=0$ carries the same mode content as the static one: the smallest Bessel zero of order $p$ exceeds $p$, which keeps the positive branch $\varepsilon=k-\Omega(m+\tfrac12)>0$ for every admissible mode when $\Omega R<1$.  This is the zero-temperature occupation invariance, stated as a theorem and verified numerically in Sec.~\ref{sec:res-t0}.  The Tolman weight of Eq.~\eqref{eq:Homega} together with the co-rotating bath defines the equilibrium ensemble on the disc.

\section{Theory}
\label{sec:theory}

Dynamical chiral symmetry breaking in three-dimensional electrodynamics is a nonperturbative large-$N$ phenomenon: once the gauge-boson propagator is dressed by the one-loop polarization, the mass function satisfies a self-consistent integral equation whose nonlinear solutions mark the broken phase~\cite{Pisarski1984}.  The same structure reappears when the fermions couple to a Chern--Simons term, where the monopole content of the vacuum becomes explicit~\cite{KondoMaris1995,KogutWang1996}, and when a four-Fermi interaction is added to noncompact QED$_3$~\cite{ArmourKogutStrouthos2010}.  On the Schwinger--Dyson side the closure has been tested against the gauge choice~\cite{BashirPennington1994,HawesWilliamsRoberts1996}, against the regularization scheme~\cite{GusyninSchreiberSizer1999}, and in the presence of an external magnetic field~\cite{LeungNgAckley1996}; for fermions confined to a brane in reduced QED, with a massive gauge boson, the breaking is studied in the same truncation~\cite{GorbarGusyninMiransky2001}.

\subsection{Logarithmic kernel}
\label{sec:theory-kernel}

The rainbow-truncation Schwinger--Dyson equation for the mass function $B(p)$ of $N$ two-component fermion flavours, with the gauge-boson propagator dressed by the one-loop polarization $\Pi(q)=Ne^2|q|/8$, closes on the scalar projection.  The trace structure of the vector vertex eliminates the $\gamma$-odd pieces exactly, so the scalar equation closes on a pure scalar translation kernel---no angular-momentum channel decomposition survives at this order.  The flavour factor enters only through the polarization; the vertex coefficient is $e^2$, not $Ne^2$.  After the three-dimensional angular average, the gap equation reduces to the logarithmic-kernel form
\begin{equation}
  B(p)=\frac{e^2}{2\pi^2 p}\int_0^\infty\!dk\;\frac{k\,B(k)}{k^2+B^2(k)}\ln\frac{p+k+\lambda}{|p-k|+\lambda},\qquad \lambda\equiv\frac{Ne^2}{8},
  \label{eq:logkernel}
\end{equation}
which depends on the flavour number only through the single coupling $g=e^2/\lambda=8/N$.  This is the same equation as Eq.~(4) of ANW88~\cite{Appelquist1988}, term by term.  The widely quoted $N_c=32/\pi^2$ is the analytic solution of a hard-cutoff approximation to this logarithm; evaluating the kernel both ways gives a ratio of $1.307$.  The derivation of the kernel, the angular reduction, and the ANW88 term-by-term identification is given in Appendix~\ref{app:B}, Eqs.~(B1)--(B3).

The breaking direction is the standard one: $N<N_c$ breaks, $N>N_c$ restores.  Large $N$ caps the logarithm on the diagonal band $|u-v|\lesssim1$ and weakens the kernel; small $N$ strengthens it.

\emph{Why the kernel closes on a pure scalar.}  The vector vertex $\Gamma_\mu=\gamma_\mu$ at the rainbow order, and the polarization is transverse: $P^{\rm T}_{\mu\nu}=\delta_{\mu\nu}-p_\mu p_\nu/p^2$.  The trace $\mathrm{Tr}_4[\gamma_\mu\,\Sigma\,\gamma_\nu]\,P^{\rm T}_{\mu\nu}$ picks up $\delta_{\mu\nu}P^{\rm T}_{\mu\nu}=2$ from the transverse projection, while the three-$\gamma$ trace vanishes by charge conjugation.  No angular-momentum channel decomposition survives: the gap equation is a single scalar integral equation in the radial momentum, not a coupled-channel system.  This closure is what makes the QED$_3$ rainbow kernel exactly tractable---the equation reduces to a one-dimensional integral that can be solved by fixed-point iteration, and the critical coupling is a well-defined quantity.  Subleading $1/N$ and vertex corrections introduce additional channel structure beyond the present closure.

\emph{ANW88 identification.}  The kernel equation above is the same equation as Eq.~(4) of Appelquist, Nash and Wijewardhana~\cite{Appelquist1988}, term by term, at the same leading order in $1/N$ (bare vertex): the coefficient $4a/(\pi^2Np)$ with $a=Ne^2/8$ equals $e^2/(2\pi^2p)$, and the logarithm $\ln[(u+v+1)/(|u-v|+1)]$ is identical.  The hard-cutoff approximation gives $N_c=32/\pi^2\approx3.24$ (their Eq.~(5)), whereas the strict-kernel evaluation gives $2.4807$.  The corresponding values $3.2423$ and $2.4807$ yield the ratio $1.307\pm0.001$, which quantifies the cutoff dependence of the critical-flavor estimate.

\subsection{Finite-temperature two-branch displacement}
\label{sec:theory-thermal}

On the rotating disc, rotation enters the thermal occupancy through the angular-momentum projection $j_a=m_a+\tfrac12$ of each mode:
\begin{equation}
  n_a=n_F\!\Big(\frac{E_a-\Omega j_a}{T_0}\Big),\qquad E_a=\sqrt{k_a^2+\omega^2},
\end{equation}
so the $+j$ and $-j$ branches are displaced symmetrically in energy.  The thermal factor multiplying the scalar mass term is
\begin{equation}
  B(E,\Omega j,T_0)=1-n_F\!\Big(\frac{E-\Omega j}{T_0}\Big)-n_F\!\Big(\frac{E+\Omega j}{T_0}\Big)=\tfrac12\!\left[\tanh\frac{E-\Omega j}{2T_0}+\tanh\frac{E+\Omega j}{2T_0}\right],
  \label{eq:thermalfactor}
\end{equation}
with two consistency anchors: at $\Omega=0$, $B=1-2n_F(E/T_0)$ (standard thermal suppression); at $T_0\to0$ with $|\Omega j|<E$, $B\to1$ (no level crossing below the flip threshold).  The linear-in-$\Omega$ pieces cancel pairwise between the $\pm j$ branches, and the expansion is
\begin{equation}
  B=1-2n_F(x)-\Big(\frac{\Omega j}{T_0}\Big)^{\!2}n_F''(x)+O(\Omega^4),\qquad x=\frac{E}{T_0},
\end{equation}
so the leading rotation effect is quadratic in $\Omega/T_0$.  The sign of the response coefficient is fixed analytically: the closed form $n_F''(x)=e^x(e^x-1)/(1+e^x)^3$ has an odd part that is negative on the negative branch, and the paired-branch combination $\sum_a j_a^2[n_F''(E_a)-n_F''(-E_a)]$ is positive definite, so rotation-lifted occupancy suppresses the condensate---the same direction as ordinary thermal suppression.  The derivation is in Appendix~\ref{app:B}, Eqs.~(B4)--(B6).

\emph{Physical content.}  The two-branch displacement is the mathematical body of the occupation channel: rotation does not create new modes or change the spectrum at $T=0$, but it reweights the thermal occupation of the $\pm j$ branches through the angular-momentum chemical potential $\Omega J_z$ in the KMS weight.  The sign of the response is a consequence of the convexity of the Fermi function: $n_F''(x)$ is negative for $x>0$ (above the Fermi surface, where occupation is depleted) and the paired-branch sum is positive definite, so the net effect is a suppression.  This is the same direction as ordinary thermal suppression---rotation heats the condensate in the sense that it further reduces the dynamical mass---but through a different mechanism: not a uniform temperature increase, but a selective reweighting of angular-momentum-carrying modes.  The coefficient is a disc-specific quantity (it depends on $\TL$, $R\Lambda$, and the modal weights); it has no flat-space limit, because the angular-momentum projection entering the occupancy is not a well-defined quantity in translationally invariant space.

\emph{Screening-channel regularization.}  The screening coefficient $\kaphat_{\rm scr}(\eps)$ is reported together with its regularization window and its logarithmic $\eps\to0$ behavior.

\subsection{Three-channel decomposition}
\label{sec:theory-decomp}

The rotation-induced shift of the dynamical mass is decomposed by a channel-isolation test with one channel active and the others frozen in the full self-consistent iteration:
\begin{equation}
  \kaphat_{\rm tot}=\kaphat_{\rm occ}+\kaphat_{\rm scr}\cdot(\TL)^2+\kaphat_{\rm mix},
  \label{eq:decomp}
\end{equation}
where $\kaphat_{\rm occ}$ is the occupation-reweighting channel (thermal occupations shifted by $\Omega j$), $\kaphat_{\rm scr}$ is the screening channel (polarization kernel modified by the rotating occupancy), and $\kaphat_{\rm mix}$ is the residual cross term.  The polarization kernel entering $\kaphat_{\rm scr}$ is
\begin{equation}
  A(Q)=\lim_{\eps\to0}\frac{\delta\Pi_{\rm scr}^{(2)}(Q)}{h(Q)},
  \label{eq:AQ}
\end{equation}
with $W(Q)=JJ/V$ the two-point structure; the regulated value is evaluated at $\eps=0.01$ and quoted at that regularization.  The decomposition is defined at the reference parameters ($N=2$, $\TL=0.01$, $R\Lambda=40$), and its channel weights are therefore specific to these reference parameters.

\section{Results}
\label{sec:results}

\subsection{Critical coupling at zero rotation}
\label{sec:res-nc}

The critical flavor number at zero rotation, defined as the disappearance point of the nonlinear solution branch, is
\begin{equation}
  \Nc(0)=2.60 \pm 0.01
  \label{eq:Nc}
\end{equation}
where the quoted uncertainty is the endpoint-localisation error from bisection.  The strict-kernel boundary $2.481$ and the Miransky-window boundary $3.045\pm0.024$ are reported as separate diagnostic scales.  Their spread $\approx0.56$ records the dependence on the definition of the critical point.  The Miransky-type fit prefers a power-law form with $\Delta\mathrm{AIC}=60$--$109$ and is used to characterize the scaling shape.

\begin{figure}[t]
  \includegraphics[width=\linewidth]{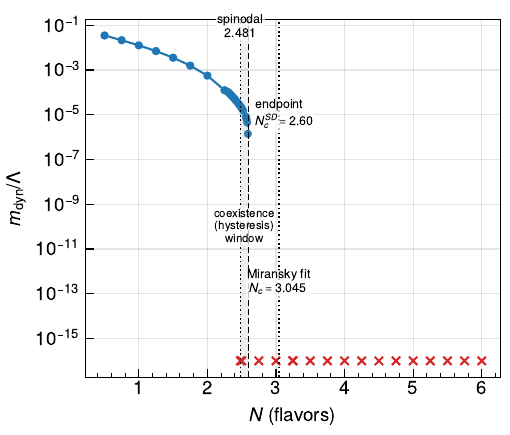}
  \caption{Dynamical mass $m_{\rm dyn}(N)$: broken branch (solid), symmetric branch (dashed), three diagnostic boundaries (strict kernel, endpoint, Miransky window).  The hysteresis window is shaded.  Solid lines: measured branch; dashed: diagnostic boundary (interpretation summarized in Table~\ref{tab:summary}).}
  \label{fig:main}
\end{figure}

The mass scaling of the dynamical mass is
\begin{equation}
  m_{\rm dyn}\propto e^2,
  \label{eq:mscale}
\end{equation}
with the numerical ratios $2.000$ and $5.012$ reproducing the analytic expectation.  The earlier literature value is reproduced as $3.2423$ when the kernel is evaluated with a hard momentum cutoff, against $2.4807$ for the strict-kernel evaluation.  Their ratio,
\begin{equation}
  \frac{N_c^{\rm hard\;cutoff}}{N_c^{\rm strict\;kernel}}=1.307\pm0.001,
  \label{eq:anw}
\end{equation}
identifies the hard-cutoff approximation as the origin of the discrepancy with the ANW88 result $N_c=32/\pi^2$~\cite{Appelquist1988}.  Figure~\ref{fig:main} shows the broken and symmetric branches, the three criteria, and the hysteresis window.

\emph{Literature context.}  The critical flavor number in QED$_3$ depends on the vertex Ansatz and the truncation scheme: Bashir \emph{et al.}~\cite{Bashir2008} showed that $N_c$ is tied to the infrared homogeneity of the photon polarization and varies with the vertex completion; Fischer \emph{et al.}~\cite{Fischer2004} found $N_f^{\rm crit}\approx4$ in an unquenched treatment with a Ball--Chiu vertex.  The rainbow truncation used here has a bare vertex, so $N_c$ is the value at this closure.  The finite-disc geometry follows the finite-volume precedent of Goecke \emph{et al.}~\cite{Goecke2009}; the $R\Lambda$ sweep reaches a $\sim3\%$ plateau, which sets the finite-cutoff systematic scale.

\emph{Dependence of the critical flavour on the implementation.}  The rainbow value quoted here is one point in a long series of implementations, and the series is informative about how much the answer depends on the prescription.  The infrared cutoff that regularizes the logarithm shifts the critical number, and the shift is quantifiable~\cite{GusyninReenders2003}; a massive gauge boson changes it as well~\cite{LiuCheng2003}, and an independent determination uses a nonlocal gauge~\cite{JiangWangLu2014}.  Beyond the rainbow level the vertex completion changes both the critical value and the infrared homogeneity that fixes it~\cite{AkramBashir2013}, while unquenched treatments with a consistently dressed photon polarization place the critical flavour of order four~\cite{Fischer2004,KizilersuSizerPennington2015}.  The equation itself has been shown to be solvable with gauge covariance preserved~\cite{JiaPenningtonPLB2017} and solved exactly in Minkowski space with on-shell renormalization~\cite{JiaPennington2017}; in reduced QED, with a massive gauge boson, the vertex structure and gauge invariance of the dynamical breaking have been re-derived at higher order~\cite{AlbinoBashirMizher2022}.  Two alternative characterizations of the transition are worth noting: the critical behaviour can be read as an annihilation of fixed points in the flow~\cite{Herbut2016}, and the parity-symmetric phase structure admits a scale-invariant formulation in which the critical flavour is located by the non-perturbative flow rather than by a fixed-point equation~\cite{KarthikNarayanan2016,GusyninPyatkovskiy2016}.  On the lattice side the critical flavour in noncompact QED$_3$ is governed by the monopole content of the ensemble~\cite{KogutWang1996,ArmourKogutStrouthos2010}, and an external magnetic field provides independent lattice evidence for the dynamical breaking~\cite{KogutSinclair2024}.

\emph{Numerical uncertainties.}  The endpoint localisation contributes $\pm0.01$, the finite-cutoff sweep reaches a $\sim3\%$ plateau, and momentum-grid discretisation contributes $0.06\%$.  The criterion-definition spread ($\sim0.56$) is reported separately from the endpoint uncertainty.  Solver convergence was checked to $10^{-12}$ ($10^{-13}$ for the full nonlinear iteration).  Bare-vertex and residual modal-truncation systematics require higher-order extensions.

\emph{Interpretation of the critical-flavor estimates.}  The three boundaries answer different questions: the strict-kernel value tests the kernel equation alone, the endpoint locates branch disappearance under the full self-consistent calculation, and the Miransky window marks where the power-law scaling form holds.  Their spread quantifies the dependence on the definition of the critical point; the three estimates are listed in Table~\ref{tab:summary}.  The endpoint ($2.60$) and strict-kernel value ($2.481$) differ by an amount comparable to the $R\Lambda$ plateau ($\sim3\%$).

\emph{Miransky scaling.}  The power-law form that the fit selects is not an artefact of the fitting window: near-critical dynamical mass generation in QED$_3$ and in closely related vector gauge theories is widely reported to follow an exponentially suppressed power law whose exponent is set by the ratio of the actual to the critical flavour, rather than by a mean-field critical exponent~\cite{BraunFischerGies2011}.  Whether a regime with only a weakly broken scale can realize the same scaling, or must be described by an analytic crossing of the fixed point, has been examined separately~\cite{AntipinDiChiaraMojaza2012}; the scaling law has also been re-derived by mapping the dynamical-mass equation to a Schr\"odinger problem with an infinite-well/barrier potential~\cite{GarciaMunozAlfaro2025}.  The window boundary reported here is the range of the flavour parameter inside which the power-law form remains preferred over the mean-field form at the stated significance; the exponent coefficient is read off from the fit to the branches and is not imposed from these works.

\subsection{Zero-temperature theorem}
\label{sec:res-t0}

Below the occupancy-flip threshold $\Omega R<1$, the total rotation response coefficient vanishes identically at zero temperature:
\begin{equation}
  \kaphat_{\rm tot}(T{=}0)\equiv 0.
  \label{eq:theorem}
\end{equation}
Both channels lose their $\Omega$-dependence in the $T_0\to0$ limit: the occupied-mode set is identical for the rotating and static problems (confirmed over $\Omega R\in\{0.3,0.5,0.8\}$ by the spectral calculation, $96$--$896$ modes per truncation), and the screening correction vanishes with the thermal factor $B\to1$.  The rotating vacuum and the static vacuum are the same state.  All rotation effects on the condensate enter through the thermal density matrix.  At $T=0$ the rotating and static density matrices, and hence the condensate, coincide.

\begin{figure}[t]
  \includegraphics[width=\linewidth]{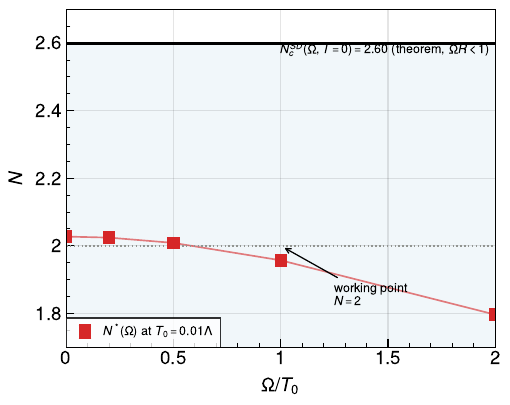}
  \caption{Phase diagram in the $(\Omega/T_0,\,N)$ plane.  Solid horizontal line at $T=0$: the theorem $\kaphat_{\rm tot}(T{=}0)\equiv0$ (constant $N_c$).  Points with error bars (localisation of the branch termination): the crossover flavor $\Nstar(\Omega)$ at fixed $T_0/\Lambda=0.01$, measured to fall monotonically.  The top axis shows $\Omega R$.  The $T=0$ line follows from the theorem; the $\Nstar(\Omega)$ curve is shown at fixed temperature.}
  \label{fig:phase}
\end{figure}

The critical flavor number is therefore unchanged by rotation at $T=0$:
\begin{equation}
  \Nc(\Omega,T{=}0)=\Nc(0).
  \label{eq:NcOmega}
\end{equation}
The theorem applies to the condensate response in the four-component Dirac formulation.  Figure~\ref{fig:phase} shows the $T=0$ constant line (theorem) and the finite-temperature crossover line.

\emph{Spectral and solver checks.}  The spectral calculation sweeps $\Omega R$ over $\{0.3,0.5,0.8\}$ and counts occupied modes at $T=0$ for truncation levels $96$, $192$, $320$, and $896$.  The rotating and static vacua have identical occupied-mode content.  The discrete-sum and exact forms of $\delta\Pi_{\rm scr}^{(2)}$ agree term by term, and the computed $\Omega$ difference is exactly zero.  Tightening the solver tolerance from $10^{-8}$ to $10^{-12}$ leaves the self-consistent solution unchanged; the first-order deformation cancels by paired parity, while the second-order coefficient is stable and nonzero ($-1.44$ with $\Omega^2$ scaling).  These checks provide numerical support for the vacuum-invariance theorem.

\subsection{Three-channel decomposition}
\label{sec:res-decomp}

At the reference parameters ($N=2$, $\TL=0.01$, $R\Lambda=40$), the decomposition of Eq.~\eqref{eq:decomp} gives:

\begin{equation}
  \kaphat_{\rm occ}=-1.5474,
  \label{eq:kocc}
\end{equation}
from finite differences, cross-checked by the analytic chain at ratio $0.98$ (order-of-magnitude and sign agreement).  Self-consistent feedback amplifies the direct effect by a factor $26.5$.  The occupancy channel accounts for $\ge99.995\%$ of the total response ($1.5474/1.547463=99.9959\%$).

\begin{equation}
  \kaphat_{\rm scr}(\eps{=}0.01)=-0.97\pm0.01,
  \label{eq:kscr}
\end{equation}
delivered at fixed regularization together with the measured band $[-1.25,-0.97]$ over $\eps\gtrsim5\times10^{-3}$ (grid floor $q_{\min}{=}10^{-4}$).  Because $A_{\rm PW}(q\lesssim10^{-4})\approx2\times10^{-6}\neq0$, the $\eps\to0$ behavior is logarithmically divergent in the present regularization.

\begin{equation}
  |\kaphat_{\rm mix}|=3.3\times10^{-5},
  \label{eq:kmix}
\end{equation}
so restricted additivity holds for these reference parameters: $\kaphat_{\rm occ}$ and $\kaphat_{\rm scr}$ may be reported separately.  The mixed residual remains a small cross-channel contribution at these parameters.

\begin{figure}[t]
  \includegraphics[width=\linewidth]{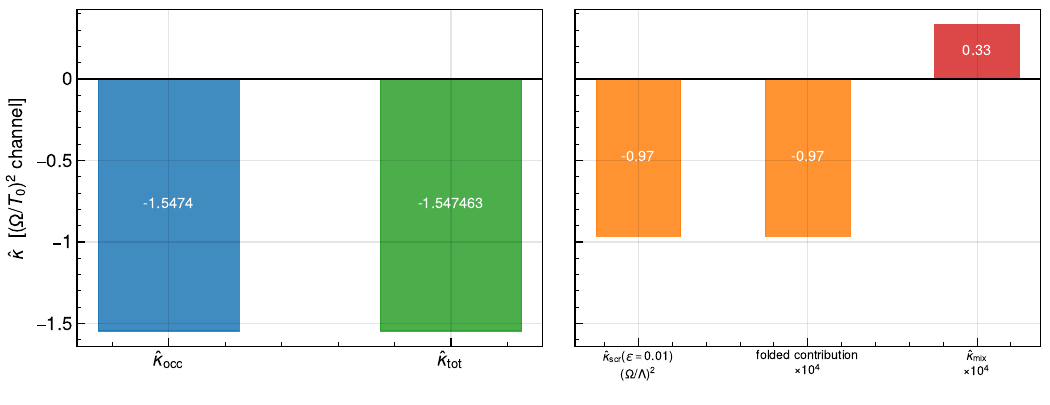}
  \caption{Three-channel decomposition.  Left: $(\Omega/T_0)^2$ scaling of the occupancy channel $\kaphat_{\rm occ}$ (solid) and total $\kaphat_{\rm tot}$ (solid).  Right: $(\Omega/\Lambda)^2$ scaling of the screening value $\kaphat_{\rm scr}(\eps{=}0.01)$ (dashed) and the mixed residual $\kaphat_{\rm mix}$ (dotted).}
  \label{fig:decomp}
\end{figure}

The full non-linear iteration gives $\kaphat_{\rm tot}=-1.547463$ (even-in-$\Omega$ residual $2\times10^{-14}$), against the linear prediction $-1.547497$.  Figure~\ref{fig:decomp} shows the two panels: $(\Omega/T_0)^2$ for the occupancy/total channel and $(\Omega/\Lambda)^2$ for the screening value and the mixed residual.

\emph{Interpretation.}  The occupancy channel dominates by about $10^4$ for these reference parameters.  The direct curvature response is set by the modal-weighted $n_F^{\prime\prime}$ sum, while self-consistent feedback $(I-J)^{-1}$ amplifies it by $26.5$.  Along a path at fixed thermal distance to $T_c$, the feedback is expected to grow as $N\to\Nc$; moving between the two parameter sets changes that thermal distance and produces the measured reduction.  The screening channel is smaller by two orders of magnitude in the reduced $(\TL)^2$ frame because the polarization correction is second order in the rotating occupancy.  Since the occupancy channel accounts for $\ge99.995\%$ of the response, the total is insensitive to the screening regularization.  The mixed residual remains a small cross-channel contribution for these parameters.

\emph{Second parameter set and the mechanism of the feedback enhancement.}  For $N{=}2.2$ and $T_0/\Lambda=8\times10^{-3}$, with a thermal distance to $T_c$ comparable to that of the reference parameters, the decomposition gives $\kaphat_{\rm occ}=-0.412$ and remains numerically stable.  The linearized kernel is dominated by a near-soft collective mode: the top-mode share is $91.6\%$ at the primary point and $77.2\%$ at the second point.  The gain potential falls from $67.0$ to $28.8$, while the ideal $(N_c-N)^{-1}$ estimate would increase; the deviation follows from the different thermal distances, finite infrared cutoff, and disc modal spectrum.  The linear chain reproduces full-iteration values to $2\%$ and $0.3\%$.  Both points lie near the transition, so the coefficients contain both $1/m_0$ transition-proximity enhancement and soft-mode amplification; the ratio $2.33$ combines both contributions and is not an isolated soft-mode measure.

\begin{figure}[t]
  \includegraphics[width=\linewidth]{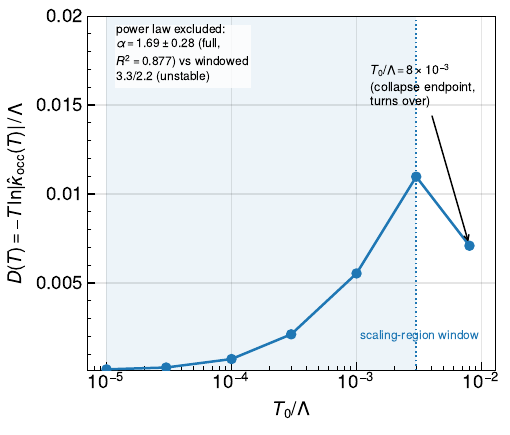}
  \caption{Effective activation temperature $D(T)$ along the symmetry-broken branch at $N{=}2.2$ (seven points).  In the scaling-region window $T_0/\Lambda\lesssim3\times10^{-3}$, $D(T)$ rises monotonically; the $8\times10^{-3}$ endpoint turns over in the transition region.  The full-window fit gives $\alpha=1.69\pm0.28$ ($R^2=0.877$), with windowed exponents $3.3\pm1.4$ and $2.2\pm0.3$.}
  \label{fig:activation}
\end{figure}

\emph{Temperature scan along the symmetry-broken branch.}  At $N{=}2.2$, scanning $T_0/\Lambda$ over seven points ($10^{-5}$ to $8\times10^{-3}$) is inconsistent with a single power law over the full window and is consistent with an activation form in the scaling-region window $T_0/\Lambda\lesssim3\times10^{-3}$.  Figure~\ref{fig:activation} shows the effective activation temperature: it rises monotonically in the scaling-region window, and the $8\times10^{-3}$ endpoint turns over in the transition region.  Below $T_0\lesssim10^{-6}\Lambda$, the lowest-mode Boltzmann weight underflows double precision, so the asymptotic saturation scale is not resolved here and is left for future asymptotic analysis.

\subsection{Crossover and temperature trend}
\label{sec:res-cross}

At fixed $T_0/\Lambda=0.01$, the crossover flavor $\Nstar(\Omega)$---the point where the broken branch terminates at finite rotation---is measured to fall monotonically over the scanned range:
\begin{equation}
  \frac{d\Nstar(\Omega)}{d\Omega}\Big|_{T_0}<0\quad\text{(measured, reference parameters)}.
  \label{eq:Nstar}
\end{equation}
This fixed-temperature trajectory is a corollary of the thermal suppression; mapping the full critical surface requires varying both $\Omega$ and $T$.

The finite-temperature crossover is directionally expected to satisfy
\begin{equation}
  T_c(\Omega)<T_c(0)\quad\text{(directional prediction)},
  \label{eq:Tc}
\end{equation}
supported by the $N{=}2$ single-point evidence (thermal suppression drives the condensate to zero at $T_c\sim0.01\Lambda$) and the pure-thermal-activation argument.  An $\Omega$-resolved scan of $T_c$ will determine its full temperature dependence.

\subsection{Numerical robustness}
\label{sec:res-consistency}

The polarization kernel $A(Q)$ satisfies its analytic identities, the universal function $I(q/T_0)$ is unchanged between the two temperature normalizations, and the $R\Lambda$ sweep reaches a $\sim3\%$ plateau.  The occupied-branch contribution is $-1.5474$ (finite difference) and $-1.547463$ (full iteration), agreeing to $10^{-4}$.

Table~\ref{tab:summary} summarizes the principal quantities and calculation parameters.

\begin{table}[t]
  \caption{Summary of principal quantities and calculation parameters.  \emph{Interpretation}: theorem, measured value, regularized value ($\eps=0.01$), corollary, or directional prediction.  Values are quoted at the reference parameters ($N{=}2$, $T_0/\Lambda=0.01$) unless stated otherwise.}
  \label{tab:summary}
  \begin{ruledtabular}
    \begin{tabular}{llll}
      Symbol                           & Value / definition       & Interpretation    & Note                             \\
      \hline
      $\Nc(0)$                         & $2.60\pm0.01$            & measured          & endpoint localisation            \\
      $\Nc(\Omega,T{=}0)$              & $=\Nc(0)$                & theorem           & Dirac formulation, $\Omega R<1$  \\
      $\kaphat_{\rm occ}$              & $-1.5474$                & measured          & $\ge99.995\%$ of linear response \\
      $\kaphat_{\rm scr}(\eps{=}0.01)$ & $-0.97\pm0.01$           & regularized value & band $[-1.25,-0.97]$             \\
      $\kaphat_{\rm mix}$              & $3.3\times10^{-5}$       & measured          & cross-channel magnitude          \\
      $\lambda_{\max}$                 & $0.985$                  & measured          & top mode $91.6\%$                \\
      $\Nstar(\Omega)$                 & monotone decrease        & corollary         & reference parameters             \\
      $T_c(\Omega)<T_c(0)$             & ---                      & prediction        & directional result               \\
      $1.307\pm0.001$                  & hard/strict kernel ratio & measured          & hard/strict cutoff ratio         \\
      \hline
      $N$                              & $2$                      & ---               & reference parameters             \\
      $\TL$                            & $0.01$                   & ---               & reference parameters             \\
      $R\Lambda$                       & $40$                     & ---               & reference parameters             \\
      $\eps$                           & $0.01$                   & ---               & regularization parameter         \\
      $q_{\min}$                       & $10^{-4}$                & ---               & grid floor                       \\
      mixing bound $\kaphat_{\rm mix}$ & $0.1$                    & ---               & calculation parameter            \\
      $\delta_{N_c}$                   & $1\%$                    & ---               & calculation parameter            \\
    \end{tabular}
  \end{ruledtabular}
\end{table}

\section{Discussion}
\label{sec:discussion}

\subsection{Mechanism}
\label{sec:disc-mech}

The sign of $\kaphat_{\rm occ}$ is fixed analytically: the paired-branch combination of $n_F^{\prime\prime}$ is positive definite, so rotation-lifted occupancy suppresses the condensate in the same direction as ordinary thermal suppression.  The magnitude is set by self-consistent feedback $(I-J)^{-1}$, which amplifies the direct curvature response by $26.5$ at the reference parameters.  The near-soft collective mode carries the amplification, with top-mode share $91.6\%$ at the primary point and $77.2\%$ at the second point.  The occupancy channel is defined by the finite-disc geometry; broader parameter scans will determine its variation.

The screening channel is a regularized value: $A_{\rm PW}(0)\approx2\times10^{-6}\neq0$ gives logarithmic $\eps\to0$ behavior.  The net leading-term cancellation in the full response chain remains an open quantitative question.

\emph{Relation to prior directions.}  The QCD literature splits on the sign of the rotation effect: continuous models~\cite{ChernodubGongyo2017,Gaspar2023} find rotation \emph{restores} chiral symmetry, while lattice studies~\cite{YangHuang2023,Nunes2025} find \emph{enhanced} breaking.  The present QED$_3$ result---rotation suppresses the condensate at finite temperature through the occupation channel---aligns with the lattice direction at the level of the sign, but the mechanism is different: the lattice enhancement is attributed to the rotation of the gauge-field background and confinement, while the QED$_3$ suppression is an angular-momentum reweighting of the thermal density matrix, with no confinement and no gauge-field rotation.  The rotating chiral soliton lattices~\cite{NishimuraYamamoto2020,Eto2024} add a third phase structure that is not present in the QED$_3$ spectrum at the reference parameters.  The QED$_3$ suppression ($\kaphat_{\rm occ}\sim-1.5$ at $(\Omega/T_0)^2\sim1$) is a thermal-occupancy effect with no QCD analogue; the QCD rotation effects are vacuum/background effects of order $\Omega/T$ in the same units.

\emph{Chiral vortical and magnetic effects.}  Rotation couples to the same anomaly sector as magnetic catalysis: for a fluid at rest in the rotating frame the angular velocity acts as a chiral chemical potential, so the anomalous-current response coefficients are fixed by vorticity and chirality together.  The chiral vortical current has been studied from the effective-action side, where the anomaly fixes its coefficient and protects it from perturbative renormalization~\cite{GolkarSon2015}, and its frequency dependence has been extracted from the current correlator~\cite{LandsteinerMegiasPenabenites2014}; the temperature dependence of the same response has been tracked separately~\cite{Kalaydzhyan2014}.  The coefficient is modified when finite rotation is combined with temperature and curvature~\cite{FlachiFukushima2018}, and the effect has been extended to fermions of arbitrary spin~\cite{HuangSadofyev2019}.  The chiral magnetic effect is the axial counterpart with vorticity replaced by a magnetic field: its current was derived in the same class of models~\cite{FukushimaKharzeevWarringa2008}, its real-time dynamics worked out~\cite{FukushimaKharzeevWarringa2010}, and the pair of effects has been cast into a common test in heavy-ion collisions~\cite{KharzeevSon2011}.  A mode decomposition of the response for rotating fermions~\cite{FukushimaShimazakiWang2020} and the chiral mass gap in curved backgrounds~\cite{FlachiFukushima2014} are the geometric complements of these results; gravitational catalysis of chiral and colour symmetry breaking in curved space belongs to the same geometric-coupling class~\cite{EbertTyukovZhukovsky2009}.  The phenomenon is not confined to gauge theories: it has been reported in Fermi liquids~\cite{KirilinKhaidukovSadofyev2012}, in photonic media~\cite{YamamotoPhotonic2017}, in relativistic and nonrelativistic Dirac systems~\cite{ShitadeMamedaHayata2020}, and, with the vorticity played by a chiral chemical potential, it catalyses dynamical chiral symmetry breaking in Dirac semimetals~\cite{BragutaKatsnelsonKotov2019}.

\emph{Magnetic catalysis and its inverse.}  The closest comparison with a rotation-dependent chiral transition is the uniform-field case in QED$_3$, where the field catalyses flavour symmetry breaking and lowers the critical flavour below its field-free value~\cite{GusyninMiranskyShovkovy1994,GusyninMiranskyShovkovy1995}.  The Schwinger--Dyson treatment in an external field is the same truncation used here~\cite{LeungNgAckley1996}; the finite-temperature version of the catalysis has been quantified~\cite{GusyninShovkovy1997}, re-derived in planar QED~\cite{CeaCosmaiGiudice2012}, and determined independently in the QED$_3$ setting~\cite{LiFengJiang2012}.  The catalysis can be reversed: thermal-magnetic effects on the effective coupling drive an inverse magnetic catalysis in which the transition temperature falls with the field~\cite{AyalaLoeweMizher2014}, and vorticity can replace the field entirely~\cite{JiangYin2022}.  The contrast with the present result is sharper still.  In a magnetic field the Landau-level spectrum is reorganized and the level density at the Fermi point grows, so the catalysis is a density-of-states effect; rotation does not reorganize the spectrum on a rigid disc, and the linear-in-$\Omega$ pieces of the two quasi-energy branches cancel pairwise, so the rotation response is carried by the occupancy of an unchanged spectrum rather than by a change of the spectrum itself.

\subsection{Experimental connection}
\label{sec:disc-exp}

\emph{Co-rotating atoms.}  Cold-atom rotation simulators~\cite{Lewenstein2007} are the direct platform: mechanical rotation at $\Omega\sim10$--$100\,$Hz is routinely achieved with $R\sim1$--$10\,\mu$m clouds and Fermi velocities $v_F\sim\,$mm/s, so the light-cylinder constraint $\Omega R<1$ is met with margin; the Feshbach-resonant tuning of the scattering length allows the effective flavor number $N$ to be swept, which is the direct handle on the crossover $\Nstar(\Omega)$.

\emph{Strained Dirac materials.}  Graphene and Dirac/Weyl semimetals realize the effective-geometry side through strain-induced pseudomagnetic fields~\cite{Jiang2017,Ren2022} and interaction-driven orders~\cite{Pikulin2016,Venderbos2016}: the pseudomagnetic field $B_{\rm ps}\sim10\,$T corresponds to a magnetic length $\ell_B=\sqrt{\hbar/eB}\sim8\,$nm, and the Coulomb-interaction regime has $\alpha_g=e^2/(4\pi\epsilon\hbar v_F)\sim2.2$~\cite{Vozmediano2010}, in the strong-coupling window where dynamical mass generation is expected.  These platforms are static (strain, not rotation): they test the effective-geometry ingredients, not the rotation response itself.  Graphene's effective flavor number $N=4$ (spin $\times$ valley) exceeds $N_c\approx2.6$; the $N=2$ parameter set is reached by restricting to a single valley or by tuning the interaction.  Curved-space Dirac prototypes---graphene wormholes and Weyl metamaterials~\cite{Gonzalez2010,Weststrom2017}---extend the geometry; whether they can host the rotating-ensemble physics (rather than only the static geometry) is an open question for the simulation platforms.  The strain-induced gauge field in Dirac/Weyl semimetals~\cite{Pikulin2016,Venderbos2016} provides the interaction-plus-geometry combination that the present framework requires; the chiral anomaly observation in strained Dirac semimetals is a qualitative confirmation that the platform accesses the relevant physics.  The crossover flavor $\Nstar(\Omega)$ is the observable most directly tied to the present results; an $\Omega$-resolved $T_c$ measurement would extend this connection.

\emph{Curved-space Dirac fermions.}  The effective-geometry ingredient on which the model rests is the low-energy Dirac description of graphene, established by two-dimensional measurements of the massless spectrum~\cite{Novoselov2005}.  Static inhomogeneity is its best-studied deformation: strain acts as a gauge field for the low-energy carriers~\cite{DeJuanManesVozmediano2013}, and an inhomogeneous strain profile enters the transport equation as a spatially varying Fermi velocity~\cite{ContrerasAstorga2020}.  Genuine curvature produces a pseudomagnetic field~\cite{CastroVillarrealRuizSanchez2017} with shifted Landau levels~\cite{DebusMendozaHerrmann2018} that have been resolved experimentally in strained samples~\cite{WagnerDeJuan2022}; local curvature and topological defects modify the electronic response in the same way as the metric~\cite{CortijoVozmediano2007}, and a curved-geometry interpretation has been proposed as the resolution of the anomalous magnetic susceptibility of curved graphene~\cite{EspinosaChampoNaumis2024}.  The geometry need not be static: time-dependent shapes generate curvature-induced pseudogauge fields~\cite{MoralesCopinger2023}.  Not every such mapping is valid, however: the curved-space Dirac description of an elastically deformed monolayer is not universally correct, and its domain of validity is set by the form of the deformation~\cite{RobertsWiseman2022}.  That restriction is the one imposed here on the finite disc: the geometry enters through the metric of the frame and the KMS ensemble, not through a deformation-induced gauge potential.

\subsection{Scope and extensions}
\label{sec:disc-lim}

The results apply to the rainbow truncation, the finite disc, and the reference parameters stated above.  Natural extensions are the rotation-dependent critical surface $\Nc(\Omega,T)$, the fixed-$N$ crossover line $T_c(\Omega)$, and the reduced-scheme subtraction family.

The $\kaphat_{\rm scr}$ value is tied to the $K_\eps$ regularization scheme; its $\eps\to0$ behavior is logarithmically divergent in the present regularization.  The $T_c(\Omega)$ inequality is a directional prediction.  Vertex completion, gauge/scheme sensitivity, angular-momentum truncation, continuum extrapolation, and independent implementation comparisons quantify the remaining systematics.

\section{Conclusion}
\label{sec:conclusion}

Within the rainbow truncation and the $\Omega R<1$ window, rotation leaves the QED$_3$ critical flavor count unchanged at zero temperature (theorem) and thermally suppresses the condensate through a dominant occupation-reweighting channel.  The critical flavor number is $\Nc(0)=2.60\pm0.01$, and the $1.307$ ratio to the hard-cutoff ANW88 value locates the result in the literature.  At the reference parameters, the occupation channel gives $\kaphat_{\rm occ}=-1.5474$ and $\ge99.995\%$ of the linearized response, the screening channel gives $\kaphat_{\rm scr}(\eps{=}0.01)=-0.97$ at fixed regularization, and $|\kaphat_{\rm mix}|=3.3\times10^{-5}$.  The crossover flavor $\Nstar(\Omega)$ falls monotonically at fixed temperature.  These results establish the finite-temperature reference at the stated parameters; the full $N_c(\Omega,T)$ surface remains to be mapped.

The quoted coefficients are specific to the stated reference parameters.  Extending the calculation to the rotation-dependent critical surface, an $T_c(\Omega)$ scan, and a lattice-regularized rotating QED$_3$ will test their parameter dependence.

\begin{acknowledgments}
  This work was supported by the Science Research Foundation for High-Level Talents of Anhui University of Science and Technology under Grant No.\ YJ20240002 and by the National Natural Science Foundation of China under Grant No.\ 11304318.  The numerical data and consistency checks supporting the results are available from the corresponding author on reasonable request.
\end{acknowledgments}

\appendix
\section{Rotating ensemble and modal basis}
\label{app:A}

\subsection{From the static to the rotating Hamiltonian}

The static flat-space Dirac Hamiltonian is $H_0=-i\alpha^i\partial_i$.  Passing to the rotating frame by the frame rotation $\Psi=e^{-i\Omega t\hat J_z}\Psi'$, with the total angular-momentum generator $\hat J_z=-i\partial_\varphi+\hat S_z$ ($\hat S_z=\tfrac{i}{2}\gamma^1\gamma^2$ in the four-component representation), turns the Dirac equation into a time-independent eigenvalue problem.  Since $[H_0,J_z]=0$ (flat-space rotation invariance),
\begin{equation}
  i\partial_t\Psi'=(H_0-\Omega\hat J_z)\Psi'\equiv H_\Omega\Psi',
  \tag{A1}
\end{equation}
and the equilibrium weight is
\begin{equation}
  \rho_{\rm eq}=\frac1Z\exp\!\big[-\beta_0(H_0-\Omega J_z)\big].
  \tag{A2}
\end{equation}
On a static eigenmode the quasi-energy is shifted $E\to E-\Omega j$ with $j=m+\tfrac12$.

\subsection{Quasi-energy spectrum and the rotating vacuum}

The single-particle quasi-energy on the disc is
\begin{equation}
  \varepsilon=s\sqrt{k_{m,n}^2+\mu_f^2}-\Omega j,\qquad k_{m,n}\;\text{fixed by the boundary family}.
  \tag{A3}
\end{equation}
This spectrum is unbounded below; the many-body ground state is the Slater determinant filling all $\varepsilon<0$ modes.  For $\Omega R<1$ the smallest Bessel zero of order $p$ exceeds $p$,
\begin{equation}
  k_{\min}^{(p)}R>p\quad\Rightarrow\quad \varepsilon=k-\Omega(m+\tfrac12)>0,
  \tag{A4}
\end{equation}
so the zero-temperature occupied-mode set is unchanged by rotation.  The occupation difference between this set and the static one ($E<0$) vanishes.

\subsection{Tolman local temperature}

The global weight is equivalent, mode by mode, to a Fermi occupation with local quantities:
\begin{equation}
  E_{\rm loc}(r)=\frac{E}{\alpha(r)},\quad \Omega_{\rm loc}(r)=\frac{\Omega}{\alpha(r)},\quad T_{\rm loc}(r)=\frac{T_0}{\alpha(r)},\quad \alpha(r)=\sqrt{1-\Omega^2r^2}.
  \tag{A5}
\end{equation}
The redshift factors cancel term-by-term: $E_{\rm loc}/T_{\rm loc}=E/T_0$ and $\Omega_{\rm loc}j/T_{\rm loc}=\Omega j/T_0$.  At the rim, $T_{\rm loc}(R)=T_0/\sqrt{1-(\Omega R)^2}$ (a factor $\approx1.67$ at $\Omega R=0.8$).
\begin{equation}
  \text{Heat bath co-rotates; }T_0\text{ defined at the disc centre.}
  \tag{A6}
\end{equation}
The redshift relation used here is the Tolman form, and it admits a proof independent of the matter content of the system in equilibrium~\cite{XiaGao2024}.  The same exponential redshift factor appears in a Rindler frame, where the thermal bath of a uniformly accelerating observer is associated with the decoherence of the local quantum states~\cite{Unruh1992}.

\subsection{Bessel reduction and the spectral condition}

The massless radial equation is Bessel's equation of order $m$, with regular solution $f(r)=J_m(kr)$.  The recursion $J_m'(z)-\frac{m}{z}J_m(z)=-J_{m+1}(z)$ gives $g(r)\propto J_{m+1}(kr)$.  Self-adjointness at $r=R$ requires
\begin{equation}
  \psi^\dagger\alpha_r\psi\big|_R=0\quad\Longleftrightarrow\quad f^*(R)g(R)\in\mathbb{R},
  \tag{A7}
\end{equation}
giving the real mixing family
\begin{equation}
  g(R)=\lambda f(R),\qquad\lambda\in\mathbb{R},
  \tag{A8}
\end{equation}
and the spectral condition
\begin{equation}
  t\,s\,J_{m+1}(kR)=\lambda\,J_m(kR).
  \tag{A9}
\end{equation}
The parity pairing $(m,+1)\leftrightarrow(-m-1,-1)$ follows from $J_{-p}=(-1)^pJ_p$.

\subsection{Zero-temperature occupation invariance}

Sweeping $\Omega R$ over $\{0.3,0.5,0.8\}$ leaves the number of occupied modes at $T=0$ unchanged (spectral calculation, $96$--$896$ modes per truncation).  The rotating vacuum at $T=0$ carries the same mode content as the static one:
\begin{equation}
  n_{\rm occ}(\Omega,T{=}0)=n_{\rm occ}(0,T{=}0).
  \tag{A10}
\end{equation}

\section{Log kernel and rotation response}
\label{app:B}

\subsection{Scalar closure and angular reduction}

The rainbow-truncation gap equation with the dressed propagator $D(q)=1/(q^2+\Pi(q))$, $\Pi(q)=Ne^2|q|/8$, closes on the scalar projection $B(p)=\mathrm{Tr}_4[\Sigma(p)]/4$.  The three-$\gamma$ trace vanishes and $\delta_{\mu\nu}P^{\rm T}_{\mu\nu}=2$, so the scalar equation closes on a pure scalar kernel:
\begin{equation}
  B(p)=\frac{2e^2}{(2\pi)^3}\int d^3k\;\frac{B(k)}{k^2+B^2(k)}\,D(|\bm p-\bm k|).
  \tag{B1}
\end{equation}
The three-dimensional angular average of the propagator is analytic,
\begin{equation}
  \int\frac{d\Omega_k}{4\pi}D(|\bm p-\bm k|)=\frac{1}{2pk}\ln\frac{p+k+\lambda}{|p-k|+\lambda},\qquad\lambda=\frac{Ne^2}{8},
  \tag{B2}
\end{equation}
giving the logarithmic-kernel equation of Eq.~\eqref{eq:logkernel}.

\subsection{Flavour-only coupling and the ANW88 identification}

In the dimensionless variables $u=p/\lambda$, $B=\lambda\psi$, the equation depends on $N$ only through $g=e^2/\lambda=8/N$:
\begin{equation}
  \psi(u)=\frac{g}{2\pi^2u}\int_0^\infty\!dv\;\frac{v\,\psi(v)}{v^2+\psi^2(v)}\ln\frac{u+v+1}{|u-v|+1}.
  \tag{B3}
\end{equation}
This is the same equation as Eq.~(4) of ANW88~\cite{Appelquist1988}: $4a/(\pi^2Np)=e^2/(2\pi^2p)$ with $a=Ne^2/8$, and the identical logarithm.

\subsection{Two-branch curvature expansion}

The thermal factor of Eq.~\eqref{eq:thermalfactor} expands as
\begin{equation}
  B=1-2n_F(x)-\Big(\frac{\Omega j}{T_0}\Big)^{\!2}n_F''(x)+O(\Omega^4),\qquad x=\frac{E}{T_0},
  \tag{B4}
\end{equation}
and the induced mass response at leading order is
\begin{equation}
  \frac{\delta M}{M}\bigg|_{\rm occ}=-\frac{1}{2T_0^2}\Big[\sum_a w_a\,j_a^2\,n_F''(E_a/T_0)\Big]\Omega^2+O(\Omega^4).
  \tag{B5}
\end{equation}
The closed form $n_F''(x)=e^x(e^x-1)/(1+e^x)^3$ has an odd part negative on the negative branch.  The paired-branch combination
\begin{equation}
  \sum_a j_a^2\big[n_F''(E_a)-n_F''(-E_a)\big]>0\quad\text{(positive definite)}
  \tag{B6}
\end{equation}
fixes the sign analytically: rotation-lifted occupancy suppresses the condensate.

\bibliographystyle{apsrev4-2}
\bibliography{ref}

\end{document}